\documentstyle[bo99,dlxtbl,epsfig_latex209,multicol]{article}

\title{X-ray Transients Monitored by the All-Sky Monitor on RXTE:  A Tabulation}
\author{Hale Bradt, Alan Levine, Ronald Remillard, and Donald A. Smith}
\affil{CSR and Physics Dept., MIT}

\def\ls{\lower 2pt \hbox{$\;\scriptscriptstyle \buildrel<\over\sim\;$}} 
\def\gs{\lower 2pt \hbox{$\;\scriptscriptstyle \buildrel>\over\sim\;$}}

\def\xref{\par\noindent\hangindent 15pt}

\begin{document}

\maketitle

\begin{abstract}
We present a tabulation of 46 transient x-ray sources monitored with
the All-Sky Monitor (ASM) on the Rossi X-ray Timing Explorer
(RXTE). They fall into four broad categories: short ($\sim1$~d),
intermediate, and long ($>500$~d) duration of outbursts, and long
period binary systems that flare up at periastron (e.g., Be
systems). The mixture of outburst/quiescent cycles and low-level
persistent emission in a few systems could indicate conditions are
near the limit for stable mass flow in the accretion disk. The two
short-time-scale systems, CI~Cam and V4641~Sgr, are within 1~kpc of the
sun, and hence many more such systems may await discovery.

\end{abstract}

\section{ASM Sky Survey}

The All-Sky Monitor (ASM; Levine et~al. 1996) on RXTE has been
monitoring the entire sky for new (uncataloged) transient x-ray
sources while also recording the intensities of the known sources. The
current catalog contains about 325 source positions of which about 180
have yielded positive detections on some occasion. The monitoring has
been reasonably continuous except for times when the sun is relatively
close to a source and except for a period of $\sim$7 weeks shortly
after launch when the detectors were turned off due to a temporary
breakdown problem. The detected sources include many well known
persistent sources as well as a substantial number of transient
sources. Some of these are recurrent and others are in their first
known outburst. Most of the latter were discovered in the RXTE era,
either with other satellites, {\it e.g.} CGRO and BSAX, or with
RXTE. Some were discovered prior to the launch of RXTE.

Of the 180 positive detections, approximately 150 reached 15~mCrab on
at least some occasion and 30 are detected at levels 2 to 15~mCrab in
averages over long periods, up to 6~months. For sources with known
positions, the detection threshold (3~sigma) away from the galactic
center is about 30 mCrab in a single sweep of the ASM cameras across
the source. A sweep usually consists of four 90-s integrations or
``snapshots'' as the cameras step across the source. The one-day
threshold (typically 5 -- 8 sweeps) can reach down to $\sim10$~mCrab.

The data are routinely searched for new ({\it i.e.,} not in the ASM
catalog) sources with a cross-correlation search of the entire
FOV. Confidence in the detection of a new persistent source arises
through multiple detections that yield crossed lines of position. In
one day, a 50-mCrab source is solidly established. Fainter sources to
about 7~mCrab can be retrieved from cross-correlation maps that
integrate one week of data. These thresholds apply to positions
reasonably removed from bright sources.

The list of detections include about 50 sources we call
``transients''. Another 23 objects are extragalactic (14~Sy1 and QSOs,
4~BL Lacs, and 5~clusters). About 40 objects exhibit periodicities in
the ASM data from the spin period of X~Per (837 s) to the 164-d
precession period of SS~433.

\section{Transient detections}

We have collected a list of the 46 brighter transients monitored with
the ASM (Table~1). We further tabulate comments about the sources in
Table~2. The criterion for inclusion on this list is that the source
be known to have been below Uhuru/HEAO-1 thresholds (few mCrab) for
sustained periods and that the source was found in a bright state of
at least 25~mCrab, as measured by the ASM.  The fainter objects omitted
include, for example, some of those detected in the galactic plane
scans with the sensitive PCA instrument on RXTE (Valinia, Kinzer, \&
Marshall 2000) or from observations with the Wide-Field Camera on
BeppoSAX (Jager et~al. 1997).

The tabulated sources are divided into several groups that depend on
the temporal character of their variability:
\begin{enumerate}
\setlength{\itemsep}{0pt}
\item two sources with very short outbursts (hours to a few days), 
\item transients of intermediate durations which have are further 
divided into 
\begin{enumerate}
\item the thirteen monitored with the ASM in the process of their
first known outburst (which may have occurred before the launch of
RXTE) and
\item nineteen that are known to be recurrent, 
\end{enumerate}
\item six sources with very long outbursts ($>$~500~d), and finally
\item six periodic systems that typically flare up when the compact
object in an elliptical orbit approaches periastron.
\end{enumerate}

The definition of a transient can be rather elusive. For example, the
existence of long-duration transients (Table~1C) suggests that there
may be no clear boundary between transients and persistent
sources. Conversely, the close binary sources X~2129+47 and
X~1755--338, long considered to be persistent sources, have
disappeared both optically and in X rays (see, {\it e.g.,} refs in van
Paradijs 1995). Neither of these sources have been detected with the
ASM to levels of a few mCrab since the Dec. 1995 launch.

\section{The tabulation}

The tabulation describes each transient in terms of the outburst
profile shape, the peak flux, the hardness ratio, the first date of
outburst, the rise and decay times and finally the duration. The light
curves exhibit much more richness than these few parameters
indicate. Sample X-ray light curves for six neutron-star systems are
shown in Fig.~1 and for six black hole systems in Fig.~2.  Plots on
expanded time-scales reveal even more detailed structure than is
evident in these figures.

\subsection{Description of data in Table 1}

{\bf Column 1:} Source name. Sources are listed in RA order within
each category. Satellite prefixes are given for objects discovered in
the past two decades, but longer-known objects are designated with the
prefix ``X''.\\
{\bf Column 2:} Type indicates black hole candidate or neutron star
system.\\
{\bf Column 3:} Outburst profiles are categorized as
fast-rise-slow-decay (frsd), symmetric, or irregular.\\
{\bf Column 4:} The peak count rate is given in mCrab. Note that
1~Crab is 75~ASM~cts/s at 1.5--12 keV.\\
{\bf Column 5:} The hardness ratio HR2 is the ratio of counting rate
in the 5 -- 12 keV band to that in the 3--5 keV band.\\
{\bf Column 6:} The start date (MJD) is the date of the first
positive detection at the onset of an outburst, or the onset of the
first outburst in Table 1D.  ``pre-XTE'' indicates the source was
first detected above threshold when RXTE observations began after
launch. MJD conversions are: \\
\makebox[5mm][l]{}1996 Jan. 0.0 = MJD 50082.0\\
\makebox[5mm][l]{}1997 Jan. 0.0 = MJD 50448.0\\
\makebox[5mm][l]{}1998 Jan. 0.0 = MJD 50813.0\\
\makebox[5mm][l]{}1999 Jan. 0.0 = MJD 51178.0\\
\makebox[5mm][l]{}2000 Jan. 0.0 =  MJD 51543.0\\
{\bf Columns 7, 8, 9:} The rise and decay times and the durations
are approximately the total time for the full rise, the exponential
time constant for the decay, and the total duration above threshold,
respectively. Some outbursts are still in progress (IP) at this date
(2000 Feb. 28).

\subsection{Description of Table 2}

The notes give descriptive features of the light curves and hardness
ratios that complement the tabulated values and also reference recent
cogent results. They are not meant to be complete; refereed
publications are favored as are later works as they ease entry into
the literature. Results from before the RXTE era may be found in the
reviews by van Paradijs~(1995) and Bradt \& McClintock~(1983).
References to earlier catalogs may also be found in these works. The
references to the table are coded based on the author and source
names.

\section{Highlights of the Tabulation}

The nature of a given source is well correlated with the ASM hardness
ratio, HR2 as follows: neutron-star low-mass binaries have HR2 =
1.0--1.5, pulsars (neutron-star high-mass binaries) have HR2 = 2--4,
and black-hole candidates exhibit large temporal variations of HR2
from extremely soft to higher values (0.3--1.5).

The outburst profiles exhibit several types of wave forms as indicated
in the table. Similarities exist from source to source and from
outburst to outburst in one source. However, there are substantial
differences also. In general, the profiles should shed light on the
disk accretion instabilities that give rise to the episodes of high
accretion luminosity.

One notable effect is the presence of long ($\sim$~1~year)
marginally-on states after a major outburst, e.g. in 1630--47 and
1608--522, and ``failed'' outbursts in Aql~X--1. These states may
indicate that the conditions for outburst are marginal. In fact,
Aql~X--1 lies on the on the thermal-viscous disk-instability boundary
(van Paradijs 1996).

The range of detected outburst durations is extremely wide as noted
above. The listed intermediate outbursts range from about $\sim 10$ to
$\sim 200$~days. The two fast x-ray novae (CI~Cam and V4641~Sgr) were
only recently discovered. These two objects are both quite close to
the sun, at distances inferred from 21 cm absorption profiles of 1.0
and 0.5~kpc respectively. It is thus possible that infrequent such
outbursts from other sources could have been missed because of the
intermittency of coverage or limited solid angles of past and present
x-ray monitoring missions. The long-duration transients are by
definition ``quasi-persistent''. These too may help reveal the factors
that lead to instability.

\section{Future work}

The ASM instrument continues to operate with most of its initial
capability, so another 1--3 years or more of useful data are
expected. The archival ASM data have recently been reprocessed with
improved {\it a posteriori} calibrations, increased temporal coverage,
and improved analysis algorithms. With these we may retrieve
additional transients. The final data base should be useful for the
determination of rates of transients, the nature of accretion
processes, and possibly may reveal new distinctions between neutron
stars and black holes.

\begin{acknowledgements}
We are grateful to the RXTE teams at MIT, GSFC, and UCSD.
\end{acknowledgements}


\begin{deluxetable}{llcrccccc}
\footnotesize
\tablecaption{RXTE ASM: Transients above 25 mCrab\label{tab:asm}}

\tablewidth{0pt}
\tablehead{
\colhead{Source} & \colhead{type} & \colhead{profile} & \colhead{peak} &
\colhead{ASM} & \colhead{start} & \colhead{Rise} & \colhead{Decay} & 
\colhead{Duration} \\
\colhead{Name} & \colhead{ } & \colhead{} & \colhead{mCrab} &
\colhead{HR2} & \colhead{date} & \colhead{days} & \colhead{days} & 
\colhead{days}
}

\startdata
\multicolumn{8}{c}{A. Fast X-ray Novae: Decay $\tau < 1$ day} \\
\\
XTE J0421+560   & bhc? & frsd &  1885 & 0.8--2.2 & 50903 & 0.3 & 0.5  & 7.7 \\
SAX J1819.3--2525 & bhc & irr  & 12200 & 0.8--2.1 & 51436 & 0.2 & 0.01 & 0.6 \\
\\
\multicolumn{8}{c}{B. Intermediate-Duration X-ray Transients (Nonperiodic)} \\
\\
\multicolumn{8}{c}{{\it Recent Initial Outbursts}} \\
XTE J0111--733  & ns &  irr? &  50 & 3.2      & 51119 & --   & -- &  53  \\
XTE J1550--564  & bhc & frsd & 6800 & 0.3--1.6 & 51062 & 4.2  & 11 & 246  \\
XTE J1723--376  & ns  & irr? &  100 & 1.5      & 51108 & --   & -- & 182  \\
GRS 1737--310   & bhc? & frsd? &  26 & 1.8      & 50497 & --   & -- &  46  \\
GRS 1739--278   & bhc & frsd &  805 & 0.6      & pre XTE &  12   & 9 &$>400$  \\
XTE J1739--285  & ??  & frsd? & 193 & 1.4      & 51471 & 6  & 22 &  $\sim$50  \\
XTE J1748--288  & bhc & frsd &  485 & 1.4      & 50966 & 1.4  & 15 &  63  \\
SAX J1750.8--2900   & ns  & frsd &  117 & 1.3      & 50515 & $<$1 & 9  &  28  \\
XTE J1755--324  & bhc & frsd &  188 & 0.3      & 50653 & 3    & 30 & 104  \\
XTE J1858+034  & ns  &  sym &   26 & 3.2      & 50842 & --   & -- &  28  \\
XTE J1859+226  & bhc & frsd & 1045 & 0.4--1.4 & 51460 & 10  & 29 & IP \\
XTE J2012+381  & bhc & frsd &  209 & 0.3      & 50956 & 3.5  & 32 & 182  \\
XTE J2123--058  & ns  & frsd &   84 & 1.2      & 50987 & $<$1 & 31 &  52  \\
\\
\multicolumn{8}{c}{{\it Known Recurrent Transients}} \\
EXO 0748--676    & ns  & frsd & $>$50 & 0.9 & pre XTE & -- & 40 &  --    \\
X 1246--588      & ns? &  sym? &  35  & 1.0 & 51271  & -- & -- & IP   \\
X 1354--644      & bhc &  sym &   52  & 1.3 & 50744  & 70 & 40 &$>$85   \\
X 1608--522      & ns  & frsd &   ??  & 1.1 & pre XTE & -- & 15 &  --    \\
   ''          &     & frsd &  911  & 1.1 & 50842  & $>8$ & 13 &  IP    \\
X 1630--472      & bhc &  irr &  336  & 1.1 & 50153  & 17 & 16 & 150    \\
   ''          &     & frsd &  416  & 1.6 & 50846  & 60 & 15 & 122    \\
   ''          &     &  irr &  215  & 1.2 & 51295  & 60 & 17 &  98    \\
GRO J1655--40    & bh  &  irr & 3138  & 0.6 & 50198  & 15 & 140 & 484    \\
X 1658--298      & ns & frsd &   45  & 0.9 & 51265  & 70 & 50 &  82  \\
\\
\multicolumn{2}{l}{{\scriptsize \it Continued Next Page.}} \\
\enddata
\end{deluxetable}

\addtocounter{table}{-1}

\begin{deluxetable}{llcrcccccl}
\footnotesize
\tablecaption{(continued) RXTE ASM: Transients above 25 mCrab}

\tablewidth{0pt}
\tablehead{
\colhead{Source} & \colhead{type} & \colhead{profile} & \colhead{peak} &
\colhead{ASM} & \colhead{start} & \colhead{Rise} & \colhead{Decay} & 
\colhead{Duration} \\
\colhead{Name} & \colhead{ } & \colhead{} & \colhead{mCrab} &
\colhead{HR2} & \colhead{date} & \colhead{days} & \colhead{days} & 
\colhead{days}
}

\startdata
\\
\multicolumn{8}{c}{{\it Known Recurrent Transients (cont.)}} \\
X 1704+241      & ns? &  sym &   33  & 1.3 & 50707  & 110 & 35 &  160  \\
RX J1709.5--266   & ns & frsd &  210  & 0.9 & 50448  & $<50$ & 50 &  86    \\
X 1711--339      & ns? &  sym &   50  & 1.2 & 51016  & 10 & 20 & 280    \\
X 1730--333  & ns  & frsd &  377  & 1.5 & mult.   & -- & -- &  25    \\
GRO J1744--28    & ns  & frsd & 1291  & 2.5 & pre XTE & -- & 65 &$>$120  \\
   ''          & ns  & frsd & 1291  & 2.5 & 50433 & 40--60 & 40 &$>$120  \\
X 1803-245      & ns? & frsd &  740  & 1.2 & 50904  & 20 & 25 &  75    \\
SAX J1808.4--3658 & ns  & frsd &  108  & 0.9 & 50333  & 8 & 8 &  19    \\
     ''        &     & frsd &   79  & 1.1 & 50911  & 4 & 12 &  21    \\
GS 1843+009     & ns  & frsd &   30  & 2.5 & 50480  & 30 & 50 & 104      \\
XTE J1856+053   & bhc &  sym &   75  & 0.4 & 50189  & 20 & 10 &  27       \\
     ''        &     & frsd &   79  & 0.4 & 50328  & 20 & 40 &  70         \\
X 1908+005        & ns  &  sym &  515  & 1.0 & mult.   & -- & -- &  78         \\
   '' (failed)          &     &  sym &   40  & 0.9 & 50232  & -- & -- & $\sim$60  \\
GS 2138+568         & ns  & sym  &   45  & 2.0 & 50630 & 12 & 10 & 27         \\
SAX J2103.5+4545 & ns  & sym  &   25  & --  & 50487 & 75 & 30 & 135  \\
   ''          &   & sym  &   25  & --  & 51471 & 12 & 12? & IP \\
\\
\multicolumn{8}{c}{C. Long Duration Transients (Duration $ > 500$ d)} \\
\\
X 1210--64      & ns? &   qp &   30  & 1.2 & pre XTE & -- & -- &  $> 675$  \\
KS J1716--389   & ?? &   qp &   50  & 1.6 & pre XTE & -- & -- & $> 1500$  \\
KS 1731--260     & ns  &   qp &  356  & 1.1 & pre XTE & -- & -- & $> 1500$  \\
GRS 1758--258    & bhc &   qp &   54  & 1.6 & pre XTE & -- & -- & $> 1500$  \\
GS 1826--238     & ns  &   qp &   35  & 1.2 & pre XTE & -- & -- & $> 1500$  \\
GRS 1915+105    & bhc & wild & 2497  & 1.3 & pre XTE & -- & -- & $> 1500$  \\
\\						    
\multicolumn{8}{c}{D. Periodic, Hard Transients} \\ 
\\						    
X 0115+634      & ns  &  sym &  400 & 3.2 & pre XTE  & -- & -- &  21 \\
RX J0812.4--3114   & ns &  sym &   25 & 2.0 & 50926 & -- & -- &  20 \\
X 1145--619      & ns  &  sym &   93 & 1.9 & 50166 & -- & -- &  30 \\
X 1845--024      & ns  &  sym &   25 & 2.5 & 50345 & -- & -- &  30 \\
X J1946+274     & ns  & mult &   80 & 2.6 & 51055 & -- & -- &  85 \\
EXO 2030+375    & ns  & mult &   25 & 1.6 & pre XTE  & -- & -- &  20 \\
\\
\enddata
\end{deluxetable}

\clearpage

\noindent
\begin{tabular}{p{12.9cm}}
\makebox[2.7cm][l]{}Table 2. Notes and References for ASM Transients\\
\\
\hline
\hline
\end{tabular}

{\footnotesize

\vspace{0.15cm}
\noindent
 {\large A. Fast X-ray Novae: Decay $\tau < 1$ day}

\vspace{0.15cm}
\noindent
 {\bf  XTE J0421+560 (CI Cam, radio jets) }

\noindent
\begin{tabular}{p{6.2cm} p{6.2cm}}
X-ray outburst (Smi99)& Radio jets (Hje99)\\
Rapid rise (few hours) and decay time scale & Optical outburst: (Wag99, Bar98)\\
\makebox[5mm][l]{}0.5 d to 2.3 d (Bel99)&  Opt spectrum (Dow84) \\
X-ray properties, unusual spectrum (Ued98, & IR spectrum, dense circumstellar wind \\
\makebox[5mm][l]{}Fro98, Orr98, Rev99)&  \makebox[5mm][l]{}(Cla99)\\ 
Distance 1.0 kpc (Orl00)& \\
\end{tabular}

\vspace{0.15cm}
\noindent
 {\bf  SAX J1819.3--2525 (V4641 Sgr, radio jets)  }

\noindent
\begin{tabular}{p{6.2cm} p{6.2cm}}
Discovery: Feb 99, (Int99), Sept. 99 & Radio jets (Hje99) \\
\makebox[5mm][l]{}(Smi99) &  Distance 0.5 kpc (Hje99)\\
Five brief X-ray outbursts in 6 days in & Opt counterpart (Gor90, Gre99) \\
\makebox[5mm][l]{}Sept. 99 (Smi99, Wij99, McC99)&  Optical outburst (Stu99, Gar99)\\
Rapid ~1-s variability (Wij99)& \\
\end{tabular}

\vspace{0.15cm}
\noindent
 {\large B. Intermediate-Duration Transients (Nonperiodic)}

{\large \it Recent Initial Outbursts}

\vspace{0.15cm}
\noindent
 {\bf  XTE J0111--733 (31-s pulsar in SMC) }

\noindent
\begin{tabular}{p{6.2cm} p{6.2cm}}
Pulsations (Cha98)& Optical counterpart (Isr99)\\
Hard X-ray profile and spinup (Wil98)& \\
\end{tabular}

\vspace{0.15cm}
\noindent
 {\bf  XTE J1550--564 (bright bhc transient) }

\noindent
\begin{tabular}{p{6.2cm} p{6.2cm}}
Acquired early in its rise (Smi98)& Optical counterpart K star at 2.5 kpc \\
Reached 6.8 Crab brightness in brief flare & \makebox[5mm][l]{}(Jai99, San99)\\ 
\makebox[5mm][l]{}(Rut98)&  Hard lags in X-ray QPO and broad band var. \\
Detected to 200 keV with BATSE (Wil98)& \makebox[5mm][l]{}(Cui00)\\ 
Evolution of spectra (Sob99)& Likely radio counterpart (Cam98)\\
QPO 0.05 --- 285 Hz (Cui99, Rem99, Hom99)& \\
\end{tabular}

\vspace{0.15cm}
\noindent
 {\bf  XTE J1723--376  }

\noindent
\begin{tabular}{p{6.2cm} p{6.2cm}}
X-ray outburst w. 816 Hz osc. (Mar99a)& X-ray position and Type I bursts (Mar99b)\\
\end{tabular}

\vspace{0.15cm}
\noindent
 {\bf  GRS 1737--310   }

\noindent
\begin{tabular}{p{6.2cm} p{6.2cm}}
Weak X-ray outburst: (Tru99, Mar97)& BSAX intensity and position (Hei97)\\
Similarity to Cyg X--1 spectrum (Cui97)& Spectrum and distance of 8500 pc (Ued97)\\
\end{tabular}

\vspace{0.15cm}
\noindent
 {\bf  GRS 1739--278 (radio emitter) }

\noindent
\begin{tabular}{p{6.2cm} p{6.2cm}}
Multiple X-ray sub-peaks (Asm00)& Candidate optical/IR object at radio \\
X-ray outburst; black-hole candidate & \makebox[5mm][l]{}position (Mar97)\\ 
\makebox[5mm][l]{}(Var97)&  X-ray spectra variations (Bor98)\\
Radio emission (Hje96)& 5-Hz QPO (Bor00)\\
\end{tabular}

\vspace{0.15cm}
\noindent
 {\bf  XTE J1739--285 }

\noindent
\begin{tabular}{p{6.2cm} p{6.2cm}}
X-ray outburst (Mar99)& \\
\end{tabular}

\vspace{0.15cm}
\noindent
 {\bf  XTE J1748--288 (radio jets, shock in ISM) }

\noindent
\begin{tabular}{p{6.2cm} p{6.2cm}}
Single outburst w. 2-d rise (Smi98)& Spectral and QPO evolution (Rev99)\\
Detected to 100 keV (Har98)& Transient radio with jet that shocked in \\
QPO at 0.5 and 32 Hz (Fox98)& \makebox[5mm][l]{}ISM (Hje98, Fen98)\\ 
\end{tabular}

\vspace{0.15cm}
\noindent
 {\bf  SAX J1750.8--29 }

\noindent
\begin{tabular}{p{6.2cm} p{6.2cm}}
Bursting transient (Nat99)& \\
\end{tabular}

\vspace{0.15cm}
\noindent
\begin{tabular}{p{12.9cm}}
\hline
\\
\end{tabular}

\clearpage

\noindent
\begin{tabular}{p{12.9cm}}
\makebox[2.7cm][l]{}Table 2. (continued) Notes and References for ASM 
Transients\\
\\
\hline
\hline
\end{tabular}

\vspace{0.15cm}
\noindent
 {\bf  XTE J1755--324 (extremely soft spectrum) }

\noindent
\begin{tabular}{p{6.2cm} p{6.2cm}}
Steep soft spectrum with hard component, & Hard X-ray flux (Gol99)\\
\makebox[5mm][l]{}(Rem97)&  ~\\
Temporal/spectral evol. similar to Nova & ~\\
\makebox[5mm][l]{}Muscae 1991 (Rev98)&  \\
\end{tabular}

\vspace{0.15cm}
\noindent
 {\bf  XTE J1858+034 (221-s pulsar) }

\noindent
\begin{tabular}{p{6.2cm} p{6.2cm}}
X-ray outburst, hard spectrum (Rem98)& QPO at 0.11 Hz (Pau98)\\
221-s pulsar (Tak98)& Celestial position (Mar98)\\
\end{tabular}

\vspace{0.15cm}
\noindent
 {\bf  XTE J1859+226 (radio source) }

\noindent
\begin{tabular}{p{6.2cm} p{6.2cm}}
X-ray outburst (Woo99)& Optical counterpart R = 15.1 (Gar99)\\
Oscillations from 0.5 Hz to 5.5 Hz (Mar99, & Opt IR consistent w. short period soft \\
\makebox[5mm][l]{}Dal99)&  \makebox[5mm][l]{}transients (Hyn99)\\ 
Detected to 200 keV w. variable cutoff & Possible optical orbital modulation 0.28 d \\
\makebox[5mm][l]{}(McC99, Dal99, Foc99)&  \makebox[5mm][l]{}(Uem99)\\ 
Radio outburst (Poo99)& \\
\end{tabular}

\vspace{0.15cm}
\noindent
 {\bf  XTE J2012+381 (radio source) }

\noindent
\begin{tabular}{p{6.2cm} p{6.2cm}}
X-ray outburst (Rem98)& Radio counterpart (Hje98, Poo98)\\
Hard initial spike and later becomes very & Optical counterpart (tentative) V = 21.3 \\
\makebox[5mm][l]{}soft, bhc (Asm00) &  \makebox[5mm][l]{}(Hyn99)\\ 
Ultra soft comp. w. hard tail in ASCA, bhc & ~\\
\makebox[5mm][l]{}(Whi98)&  ~\\
\end{tabular}

\vspace{0.15cm}
\noindent
 {\bf  XTE J2123--058 (high-lat. LMXB) }

\noindent
\begin{tabular}{p{6.2cm} p{6.2cm}}
High galactic latitude --36.2 (Lev98)& Optical outbursts (Gne99)\\
Atoll LMXB, bursts, twin kHz (Hom99, Tom99)& Precursor activity may be solar \\
Optical counterpart w. 6-h orbit (Tom99, & \makebox[5mm][l]{}contamination (Asm00)\\ 
\makebox[5mm][l]{}Sor99)&  \\
\end{tabular}

\vspace{0.15cm}
\noindent
 {\large \it Known Recurrent Transients}

\vspace{0.15cm}
\noindent
 {\bf  EXO 0748--676 (eclipsing LMXB) }

\noindent
\begin{tabular}{p{6.2cm} p{6.2cm}}
Soft x-ray excess (Bri97)& Quiescent properties (Gar99)\\
Eclipse Timings (Her97)& QPO 0.6 -- 2.4 Hz (Hom99)\\
Progressive covering of disk corona (Chu98)& \\
\end{tabular}

\vspace{0.15cm}
\noindent
 {\bf  X 1246--588     }

\noindent
\begin{tabular}{p{6.2cm} p{6.2cm}}
Probable X-ray Burster (Pir97)& Probable ROSAT source \\
~& \makebox[5mm][l]{}1RXS~J124938.0--590525~(Bol97)\\ 
\end{tabular}

\vspace{0.15cm}
\noindent
 {\bf  X 1354--644  (LMXB, BW Cir) }

\noindent
\begin{tabular}{p{6.2cm} p{6.2cm}}
Modest  outburst (Rem97)& Ginga detection (Kit90)\\
Detected to 200 keV (Har97)& Low/hard state; rapid var. (Rev00)\\
\end{tabular}

\vspace{0.15cm}
\noindent
 {\bf  X 1608--522 (bright recurrent LMXB) }

\noindent
\begin{tabular}{p{6.2cm} p{6.2cm}}
Sustained one-year low states after each & KHz QPO peak separation not constant \\
\makebox[5mm][l]{}outburst (Asm00) &  \makebox[5mm][l]{}(Men99b)\\ 
KHz QPO (Men98)& Island state kHz QPO (Yu97)\\
KHz QPO freq. dependence on position in & Outburst with hard spectrum (Zha96)\\
\makebox[5mm][l]{}color diagram (Men99a)&  ~\\
Quiescent luminosity (Asa96) possibly & ~\\
\makebox[5mm][l]{}thermal (Rut99)&  \\
\end{tabular}

\vspace{0.15cm}
\noindent
\begin{tabular}{p{12.9cm}}
\hline
\\
\end{tabular}

\clearpage

\noindent
\begin{tabular}{p{12.9cm}}
\makebox[2.7cm][l]{}Table 2. (continued) Notes and References for ASM 
Transients\\
\\
\hline
\hline
\end{tabular}

\vspace{0.15cm}
\noindent
 {\bf  X 1630--472 (bright recurrent bhc, 184 Hz, radio source)  }

\noindent
\begin{tabular}{p{6.2cm} p{6.2cm}}
Three outbursts w. intervals of ~700 d and & Historical outbursts behavior (Kuu97)\\
\makebox[5mm][l]{}~ 450 d (Asm00)&  Absorption dips (Kuu98)\\
Double-peaked and flat-topped profiles & Evolution of spectral components (Oos98)\\
\makebox[5mm][l]{}(Asm00)&  QPOs 0.06 -- 14 Hz (Die00)\\
Sustained (1 year) low state after 2nd & QPO 184 Hz (Rem99)\\
\makebox[5mm][l]{}outburst (Asm00)&  Radio and Hard X-rays (Hje99)\\
\end{tabular}

\vspace{0.15cm}
\noindent
 {\bf  GRO J1655--40 (rel. radio jets; 300 Hz QPO) }

\noindent
\begin{tabular}{p{6.2cm} p{6.2cm}}
Black hole, radio jets, ``microquasar'' & Low freq. QPOs; 300 Hz when source hard \\
\makebox[5mm][l]{}(Tin95)&  \makebox[5mm][l]{}(Rem99)\\ 
Mass   6 -- 7 M$_\odot$ (Oro97a, Sha99)& Spectral evolution (Men98, Tom99, Sob99)\\
Optical turn-on precedes X-ray by & Echo mapping (X-ray to optical) (Hyn98)\\
\makebox[5mm][l]{}5~d (Oro97b)&  \\
\end{tabular}

\vspace{0.15cm}
\noindent
 {\bf  X 1658--298  (X-ray burster)   }

\noindent
\begin{tabular}{p{6.2cm} p{6.2cm}}
Recovery by BSAX and X-ray burst (Hei99) & \\
\end{tabular}

\vspace{0.15cm}
\noindent
 {\bf  X 1704+241 (HD 154791) }

\noindent
\begin{tabular}{p{6.2cm} p{6.2cm}}
Peculiarities in M Giant spectrum (Gau99)& \\
\end{tabular}

\vspace{0.15cm}
\noindent
 {\bf  RX J17095--26  }

\noindent
\begin{tabular}{p{6.2cm} p{6.2cm}}
Hard X-ray outburst (Mar97)& X-ray bursts (Coc98)\\
Possible radio counterpart (Hje97)& \\
\end{tabular}

\vspace{0.15cm}
\noindent
 {\bf  X 1711--339     }

\noindent
\begin{tabular}{p{6.2cm} p{6.2cm}}
Recovered Ariel-5 and SAS-3 source (Rem98)& \\
\end{tabular}

\vspace{0.15cm}
\noindent
 {\bf  X 1730--333 (Rapid Burster)  }

\noindent
\begin{tabular}{p{6.2cm} p{6.2cm}}
Seven outbursts w. intervals ~210 d (Asm00)& followed by Type II accretion bursts \\
Outbursts last ~ 5 weeks w. two phases: & \makebox[5mm][l]{}(Gue99)\\ 
\makebox[5mm][l]{}Type I thermonuclear bursts&  \\
\end{tabular}

\vspace{0.15cm}
\noindent
 {\bf  GRO J1744--28 (bursting pulsar) }

\noindent
\begin{tabular}{p{6.2cm} p{6.2cm}}
Hard X-ray pulsations 0.47 s (Fin96) & Propeller effect (Cui97)\\
Pulsar phase changes associated with bursts & HEXE/Mir-Kvant observations (Bor97)\\
\makebox[5mm][l]{}(Kos98 and refs therein)&  Hard X-ray bursts with Konus and Mir-
Kvant \\
QPOs (Zha96, Kom97)& \makebox[5mm][l]{}(Apt98, Ale98)\\ 
Super-Eddington fluxes imply beaming & X-ray properties from BATSE and ASCA \\
\makebox[5mm][l]{}(Gil96)&  \makebox[5mm][l]{}(Woo99, Nis99)\\ 
Possible near IR counterpart (Aug97)& \\
\end{tabular}

\vspace{0.15cm}
\noindent
 {\bf  X 1803--245  (XTE J1806--246)   }

\noindent
\begin{tabular}{p{6.2cm} p{6.2cm}}
X-ray outburst (Mar98)& Possible burst source (Mul98) \\
QPOs (Rev99, Wij99)& \\
\end{tabular}

\vspace{0.15cm}
\noindent
 {\bf  SAX J1808.4--36 (401-Hz accreting pulsar) }

\noindent
\begin{tabular}{p{6.2cm} p{6.2cm}}
X-ray outburst (Int98)& Broad-band power spectrum (Wij98b)\\
401-Hz pulsations and 2-h orbit (Wij98a, & Optical counterpart (Roc98)\\
\makebox[5mm][l]{}Cha98)&  Transient radio emission (Gae99)\\
Soft phase lags (Cui98, Vau98)& Renewed activity Feb. 00 (vdK00, Wac00)\\
X-ray spectrum (Gil98, Hei98)& \\
\end{tabular}

\vspace{0.15cm}
\noindent
\begin{tabular}{p{12.9cm}}
\hline
\\
\end{tabular}

\clearpage

\noindent
\begin{tabular}{p{12.9cm}}
\makebox[2.7cm][l]{}Table 2. (continued) Notes and References for ASM 
Transients\\
\\
\hline
\hline
\end{tabular}

\vspace{0.15cm}
\noindent
 {\bf  GS 1843+009 (30-s pulsar) }

\noindent
\begin{tabular}{p{6.2cm} p{6.2cm}}
100-d flare followed by weak activity & X-ray recovery, 30-s pulse period, and \\
\makebox[5mm][l]{}(Asm00)&  \makebox[5mm][l]{}spectrum (Wil97, Tak97)\\ 
\end{tabular}

\vspace{0.15cm}
\noindent
 {\bf  XTE J1856+053 (bhc) }

\noindent
\begin{tabular}{p{6.2cm} p{6.2cm}}
X-ray outburst (Mar96)& Hard X-ray flux (Bar96) \\
Soft spectrum (Asm00)& \\ 
\end{tabular}

\vspace{0.15cm}
\noindent
 {\bf  X 1908+005 (Aql X--1; bright recurrent transient) }

\noindent
\begin{tabular}{p{6.2cm} p{6.2cm}}
Five strong and two failed outbursts & Low-energy phase lags (For99)\\
\makebox[5mm][l]{}(Asm00)&  Propeller effect (Cam98)\\
Optical counterpart clarified: V = 21.6, & Outside-in outburst (opt-IR-X-ray) (Sha98) \\
\makebox[5mm][l]{}late K (Che99)&  At thermal-viscous instability boundary \\
KHz oscillations change freq after burst & \makebox[5mm][l]{}(vaP96)\\ 
\makebox[5mm][l]{}(Yu99)&  \\
\end{tabular}

\vspace{0.15cm}
\noindent
 {\bf  GS 2138+568 (Cep X--4(?), 66-s pulsar) }

\noindent
\begin{tabular}{p{6.2cm} p{6.2cm}}
Be star optical counterpart: (Bon98)& X-ray pulse profile changes (Muk00)\\
Spindown rate (Wil99)& \\
\end{tabular}

\vspace{0.15cm}
\noindent
 {\bf  SAX J2103.5+4545 (359-s pulsar) }

\noindent
\begin{tabular}{p{6.2cm} p{6.2cm}}
Faint transient, 359-s pulsar (Hul98)& Second outburst (Bay00)\\
\end{tabular}

\vspace{0.15cm}
\noindent
 {\large C. Long Duration Transients (Duration $ > 500$ d)}

\vspace{0.15cm}
\noindent
 {\bf  X 1210--64  (quasi-persistent)   }

\noindent
\begin{tabular}{p{6.2cm} p{6.2cm}}
On until 50763 (Asm00)& Uhuru and OSO--7 source\\
\end{tabular}

\vspace{0.15cm}
\noindent
 {\bf  KS J1716--389  (100-d dipper) }

\noindent
\begin{tabular}{p{6.2cm} p{6.2cm}}
Galactic center source (Ale95)& Periodicity $\sim100$~d (Wen99)\\
On until MJD 50763 (Asm00)& ~\\
Quasi persistent source with periodic dips & ~\\
\makebox[5mm][l]{}(Rem99)&  \\
\end{tabular}

\vspace{0.15cm}
\noindent
 {\bf  KS 1731--260  (524 Hz during bursts)  }

\noindent
\begin{tabular}{p{6.2cm} p{6.2cm}}
KHz QPO at 524 Hz during burst (Smi97)& ROSAT observations, celestial position, \\
Two KHz QPO at 898 and 1159 Hz (Wij97)& \makebox[5mm][l]{}persistent source? 
(Bar98)\\ 
\end{tabular}

\vspace{0.15cm}
\noindent
 {\bf  GRS 1758--258 (bright hard galactic center source)}

\noindent
\begin{tabular}{p{6.2cm} p{6.2cm}}
Radio jets, x-ray spectral var. and & Optical candidates (Mar98)\\
\makebox[5mm][l]{}similarity to 1E~1740.7--2942 (Smi97) & Long-term monitoring (Mai99, Kuz99) \\
ASCA spectrum, soft excess (Mer97) & \\
\end{tabular}

\vspace{0.15cm}
\noindent
 {\bf  GS 1826--238 (burster)  }

\noindent
\begin{tabular}{p{6.2cm} p{6.2cm}}
Bursts at reg. intervals (Ube97)& Possibly steady accretor since 1988 (Ube97, \\
Spectrum and distance from bursts (Int99)& \makebox[5mm][l]{}Int99)\\ 
\end{tabular}

\vspace{0.15cm}
\noindent
 {\bf  GRS 1915+105 (microquasar)  }

\noindent
\begin{tabular}{p{6.2cm} p{6.2cm}}
Superluminal jets (Mir94, Fen99)& Disk emptying episodes (Bel97, Poo97, \\
Ten distinct accretion states, some & \makebox[5mm][l]{}Fer99)\\ 
\makebox[5mm][l]{}oscillatory (Gre97, Mun99)&  Hard phase lags for 67 Hz QPO (Cui99)\\
Variable low freq. QPOs and persistent 67 & Interplay between QPOs and spectral \\
\makebox[5mm][l]{}Hz when spectrum hard (Mor97)&  \makebox[5mm][l]{}components 
(Mun99, Mar99, Fer99)\\ 
Coincident X-ray, IR and radio outbursts & ~\\
\makebox[5mm][l]{}(Poo97, Eik98, Mir98)&  ~\\
\end{tabular}

\vspace{0.15cm}
\noindent
\begin{tabular}{p{12.9cm}}
\hline
\\
\end{tabular}

\clearpage

\noindent
\begin{tabular}{p{12.9cm}}
\makebox[2.7cm][l]{}Table 2. (continued) Notes and References for ASM 
Transients\\
\\
\hline
\hline
\end{tabular}

\vspace{0.15cm}
\noindent
 {\large D. Periodic, Hard Transients}

\vspace{0.15cm}
\noindent
 {\bf  X 0115+634 (P = 24-d, 3.61 s) }

\noindent
\begin{tabular}{p{6.2cm} p{6.2cm}}
$\sim8$ maxima detected through 2/00 (Asm00)& \makebox[5mm][l]{}24-d orbital period 
(Asm00)\\ 
Major outburst Feb. 99 (Asm00) & Four cyclotron lines, (Hei99, San99)\\
Mini outbursts May -- July 96 at multiples of & Optical counterpart reclassified (Ung99) \\
\end{tabular}

\vspace{0.15cm}
\noindent
 {\bf  RX J08124--3114 (P = 81.4 d, 32 s) }

\noindent
\begin{tabular}{p{6.2cm} p{6.2cm}}
$\sim7$ maxima detected through 2/00 (Asm00)& Orbital period ~80 d (Cor00)\\
Optical counterpart is Be star LS992 & X-ray pulsar 31.9 s (Rei99)\\
\makebox[5mm][l]{}(Mot97)&  \\
\end{tabular}

\vspace{0.15cm}
\noindent
 {\bf  X 1145--619 (P = 189 d, 292 s)  }

\noindent
\begin{tabular}{p{6.2cm} p{6.2cm}}
$\sim4$ maxima detected through 2/00 (Asm00)& Multiwavelength observations, 13-yr 
review \\
Outburst (Cor96)& \makebox[5mm][l]{}(Ste97)\\ 
\end{tabular}

\vspace{0.15cm}
\noindent
 {\bf  X 1845--024 (P = 242 d, 95 s) }

\noindent
\begin{tabular}{p{6.2cm} p{6.2cm}}
(= GRO J1849--03 = GS1843--02)& BATSE outbursts w. 242-d period (Fin99)\\
$\sim$5 maxima detected through 2/00 & GRO source identified as \\
\makebox[5mm][l]{}(Asm00)&  \makebox[5mm][l]{}Ariel--5/SAS--3/Ginga source (Sof98, 
Fin99)\\ 
\end{tabular}

\vspace{0.15cm}
\noindent
 {\bf  X J1946+274 (= 3A 1942+274; P = 80 d, 16-s) }

\noindent
\begin{tabular}{p{6.2cm} p{6.2cm}}
$\sim7$ maxima detected through 2/00 (Asm00)& Pulsar, P = 16 s, (Smi98)\\
First detections since 1976 (Asm00)& Orbital period 80 d, (Cam99)\\
\end{tabular}

\vspace{0.15cm}
\noindent
 {\bf  EXO 2030+375 (P = 46 d, 42 s) }

\noindent
\begin{tabular}{p{6.2cm} p{6.2cm}}
$\sim30$ maxima detected through 2/00 (Asm00)& Thirteen outbursts at 46-d intervals, orbit 
\\
Pulse period dependence on luminosity & \makebox[5mm][l]{}from pulse phases (Sto99)\\ 
\makebox[5mm][l]{}(Rey96)&  Spectra at low luminosities (Rei99)\\
Timing properties (Rei98a)& ~\\
Long-term variability and IR spectroscopy & ~\\
\makebox[5mm][l]{}(Rei98b)&  \\
\end{tabular}

\vspace{0.3cm}

\noindent
{\Large \sc References For Table 2}

\begin{multicols}{2}

{\footnotesize 
\xref Ale95\_1716: Aleksandrovich, N. L., Aref'ev, V. A., Borozdin, K. N., Sunyaev, R. A., \& 
Skinner, G. K. 1998, Astron. 
Letters, 21, 431
\xref Ale98\_1744: Aleksandrovich, N. L., Borozdin, K. N., Aref'ev, V. A., Sunyaev, R. A., \& 
Skinner, G. K. 1998, Astron. 
Letters, 24, 7
\xref Apt98\_1744: Aptekar', R. L., et al. 1998, ApJ, 493, 404 
\xref Asa96\_1608 Asai, K., Dotani, T., Mitsuda, K., Hoshi, R., Vaughan, B., Tanaka, Y., \& 
Inoue, H. 1996, PASJ, 48, 257
\xref Asm00\_xxxx: The ASM Teams at MIT and GSFC, public data.
\xref Aug97\_1744: Augusteijn, T., et al. 1997, ApJ, 486, 1013
\xref Bar96\_1856: Barret, D., et al. 1996, IAUC 6519
\xref Bar98\_1731: Barret, D., Motch, C., \& Predehl, P. 1998, A\&A, 329, 965
\xref Bar98\_0421: Barsukova, E.A., et al. 1998, Bull. Spec. Astrophys. Obs., 45, 14 (astro-
ph/9905338)
\xref Bay00\_2103: Baykal, A., Stark, M., Swank, J. H 2000, IAUC 7355
\xref Bel97\_1915: Belloni, T., Mendez, M., King, A. R., van der Klis, M., \& van Paradijs, J. 
1997, ApJL, 488, L109
\xref Bel99\_0421: Belloni, T., et al. 1999, ApJ, 527, 345
\xref Bol97\_1246: Boller, T., Haberl, F., Voges, W., Piro, L., \& Heise, J. 1997, IAUC 6546
\xref Bon98\_2138: Bonnet-Bidaud, J. M., \& Mouchet, M. 1998, A\&A, 332, L9

}
\end{multicols}

\noindent
\begin{tabular}{p{12.9cm}}
\hline
\\
\end{tabular}

\clearpage

\noindent
\begin{tabular}{p{12.9cm}}
\makebox[2.7cm][l]{}Table 2. (continued) Notes and References for ASM 
Transients\\
\\
\hline
\hline
\end{tabular}
\begin{multicols}{2}

{\footnotesize 
\xref Bor97\_1744: Borkus, V. V., et al. 1997, Astron. Letters, 23, 421
\xref Bor98\_1739--278: Borozdin, K. N., Revnivtsev, M. G., Trudolyubov, S. P., 
Aleksandrovich, N. L., Sunyaev, R. A., \& Skinner, G. K. 1998, Astron. Letters, 24, 435
\xref Bor00\_1739--278: Borozdin, K. N. \& Trudolyubov, S. P. 2000, ApJL (submitted), astroph/9911290
\xref Bri97\_0748: Brian, T, Corbet, R., Smale, A., Asai, K.,  \& Dotani, T. 1997, ApJL, 480, L21
\xref Cam98\_1550: Campbell-Wilson, D., et al. 1998 IAUC 7010
\xref Cam98\_1908: Campana, S., et al. 1998, ApJL, 499, L65
\xref Cam99\_1946: Campana, S. Israel, G., \& Stella, L. 1999, A\&A, 352, L91
\xref Cha98\_0111: Chakrabarty, D., Levine, A. M., Clark, G. W., \& Takeshima, T. 1998, 
IAUC 7048
\xref Cha98\_1808: Chakrabarty, D., \& Morgan, E. H. 1998, Nature 394, 346
\xref Cla99\_0421: Clark, J. S., Steele, I. A., Fender, R. P., Coe, M. J. 1999, A\&A, 348, 888
\xref Che99\_1908: Chevalier, C., Ilovaisky, S. Leisy, P., \& Patat, F. 1999, A\&A, 347, L51
\xref Chu98\_0748: Church, M. J., Balucinska-Church, M., Dotani, T., \& Asai, K. 1998, ApJ, 
504, 516
\xref Coc98\_1709: Cocchi, M., et al., ApJL, 508, L163
\xref Cor96\_1145--619: Corbet, R., \& Remillard, R. 1996, IAUC 6486
\xref Cor00\_0812: Corbet, R., \& Peele, A. G. 2000, ApJL, 530, L33
\xref Cui97\_1737: Cui, Wei., Heindl, W. A.,  Swank, J. H., Smith, D. M., Morgan, E. H., 
Remillard, R., \& Marshall, F. E. 
1997, ApJL, 487, L73
\xref Cui97\_1744: Cui, W. 1997, ApJL, 482, L163
\xref Cui98\_1808: Cui, W., Morgan, E. H., \& Titarchuk, L. G. 1998, ApJL, 504, L27
\xref Cui99\_1550: Cui, E., Zhang, S. N., Chen, W., \& Morgan, E. H. 1999, ApJL, 512, L43
\xref Cui99\_1915: Cui, W. 1999, ApJL, 524, L59
\xref Cui00\_1550: Cui, W., Zhang, \& S. N., Chen, W. 2000, ApJL, 531, L45
\xref Dal99\_1859+226: Dal Fiume, D., et al. 1999, IAUC 7291
\xref Die00\_1630: Dieters et al. 2000, ApJ, in press (astro-ph 991202)
\xref Dow84\_0421: Downes, R. 1984, PASP, 96, 80
\xref Eik98\_1915: Eikenberry, S.S., Matthews, K., Morgan, E. H., Remillard, R. A., \& Nelson, 
R. W. 1998, ApJL, 494, L61
\xref Fen98\_1748: Fender, R. P., \& Stappers, B. W. 1998, IAUC 6937
\xref Fen99\_1915: Fender, R. P., et al. 1999, MNRAS, 304, 865
\xref Fer99\_1915: Feroci, M., Matt, G., Pooley, G., Costa, E., Tavani, M., \& Belloni, T. 1999, 
A\&A, 351, 985
\xref Fin96\_1744: Finger, M. H., Koh, d. T., Nelson, R. W., Prince, T. A., Vaughan, B. A., \& 
Wilson, R. B. 1996, Nature, 
381, 291
\xref Fin99\_1845: Finger, M. H., et al. 1999, ApJ, 517, 449
\xref Foc99\_1859+226: Focke, W. B., Markwardt, C. B., Swank, J. H., \& Taam, R. E. 1999, 
BAAS, 31, 1555
\xref For99\_1908: Ford, E. C. 1999, ApJL, 519, L73
\xref Fox98\_1748: Fox, D., \& Lewin, W. 1998, IAUC 6964
\xref Fro98\_0421: Frontera, F., et al.1998 A\&A, 339, L69
\xref Gae99\_1808: Gaensler, B. M., Stappers, B. W., \& Getts, T. J. 1999, ApJL, 522, L117
\xref Gar99\_0748: Garcia, M. R., \& Callanan, P. J. 1999, AJ, 118, 1390
\xref Gar99\_1819: Garcia, M.R., \& McClintock, J.E. 1999, IAUC 7271
\xref Gar99\_1859+226: Garnavich, P, M., Stanek, K. Z., \& Berlind, P.  1999, IAUC 7276
\xref Gau99\_1704: Gaudenzi, S., \& Polcari, V. F. 1999, A\&A, 347, 4
\xref Gol99\_1755: Goldoni, P., et al. 1999, ApJ, 511, 847
\xref Gor90\_1819: Goranskij 1990, IBVS 346
\xref Gil96\_1744: Giles, A. B., Swank, J. H., Jahoda, K., Zhang, W., Strohmayer, T., Stark, M. 
J., \& Morgan, E. H 1996, 
ApJL, 469, L25
\xref Gil98\_1808: Gilfanov, M., Revnivtsev, M., Sunyaev, R., \& Churazov, E. 1998, A\&A, 
338, L83
\xref Gne99\_2123: Gneiding, C. D., Steiner, J. E., \& Cieslinski, D. 1999, A\&A, 352, 543

}
\end{multicols}

\noindent
\begin{tabular}{p{12.9cm}}
\hline
\\
\end{tabular}

\clearpage

\noindent
\begin{tabular}{p{12.9cm}}
\makebox[2.7cm][l]{}Table 2. (continued) Notes and References for ASM 
Transients\\
\\
\hline
\hline
\end{tabular}
\begin{multicols}{2}

{\footnotesize 
\xref Gre97\_1915: Greiner, J., Morgan, E. H., \& Remillard, R. A. 1997, ApJL, 473, L79
\xref Gre99\_1819: Green, D.W.E. 1999, IAUC 7277
\xref Gue99\_1730: Guerriero, R., et al. 1999, MNRAS, 307, 179
\xref Har97\_1354: Harmon, B. A., \& Robinson, C. R. 1997, IAUC 6774
\xref Har98\_1748: Harmon, B. A., McCollough, M. L., Wilson, C. AS., Zhang, S. N., \& Paciesas, W. S. 1998, IAUC 6933
\xref Hei97\_1737: Heise, J. IAUC 6606
\xref Hei98\_1808: Heindl, W. A., \& Smith, D. M. 1998, ApJL, 506, L35
\xref Hei99\_0115: Heindl, W. A., et al. 1999, ApJL, 521, L49
\xref Hei99\_1658: Heise, J., et al. 1999, IAUC 7263
\xref Her97\_0748: Hertz, P., Wood, K., \& Cominsky, L. 1997, ApJ, 486, 1000
\xref Hje96\_1739--278: Hjellming, R. M., Rupen, M. P., Marti, J., Mirabel, F., \& Rodriguez, L. 
F. IAUC 6383
\xref Hje97\_1709: Hjellming, R. M., \& Rupen, M. P. 1997, IAUC 6547
\xref Hje98\_1748: Hjellming, R. M., et al. 1998, Paper in preparation; see Abstract \#103.08, 
AAS Mtg \#193
\xref Hje98\_2012: Hjellming, R. M., \& Rupen, M. P. 1998, IAUC 6924; see also IAUC. 6932
\xref Hje99\_0421: Hjellming, R.M., \& Mioduszewski, A.J. 1999, IAUC 6862
\xref Hje99\_1608: Hjellming, R. M., et al. 1999, ApJ, 514, 383
\xref Hje99\_1819: Hjellming., R.M., et al. 1999, IAUC 7265; see also 2000, BAAS, Meeting 
195, late papers
\xref Hom99\_0748: Homan, J., Jonker, P. G., Wijnands, R., van der Klis, M., \& van Paradijs, J. 
1999, ApJL, 516, L91
\xref Hom99\_1550: Homan, J., Wijnands, R., \& van der Klis 1999, IAUC 7121
\xref Hom99\_2123: Homan, J., Mendez, M., Wijnands, R., van der Klis, M., \& van Paradijs, J. 
1999, ApJL, 513, L119
\xref Hul98\_2103: Hulleman, F., In't Zand, J. J. M., \& Heise, J. 1998, A\&A, 337, L25
\xref Hyn98\_1655: Hynes, R. I., O'Brien, K., Horne, K., Chen, W., \& Haswell, C. A. 1998, 
MNRAS, 299, L37
\xref Hyn99\_1859+226: Hynes, R. I., Haswell, A. J., \& Chaty, S. 1999, IAUC 7294
\xref Hyn99\_2012: Hynes, R. I., Roche, P., Charles, P. A., \& Coe, M. J. 1999, MNRAS, 305, 
L49
\xref Int98\_1808: In 't Zand, J. J. M., Heise, J., Muller, J. M., Bazzano, A., Cocchi, M., 
Natalucci, L., Ubertini, P. 1998, 
A\&A, 331, L25
\xref Int99\_1819  In't Zand, J., et al. 1999, IAUC 7119
\xref Int99\_1826: In 't Zand, J. J. M., Heise, J., Kuulkers, E., Bazzano, A., Cocchi, M., Ubertini, 
P. 1999, A\&A, 347, 891
\xref Isr99\_0111: Israel, G. L., Stella, L., \& Mereghetti, S. 1999, IAUC 7101
\xref Jai99\_1550: Jain, R. K., Bailyn, C. D., Orosz, J. A., Remillard, R. A., \& McClintock, J. E. 1999, ApJL, 517, L131
\xref Kit90\_1354: Kitamoto, S., et al. 1990, ApJ, 361, 590
\xref Kom97\_1744: Kommers, J. M., Fox, D. W., Lewin, W. H. G., Rutledge, R. E., van Paradijs, J., \& Kouveliotou, C. 1997, ApJL, 482, L53
\xref Kos98\_1744: Koshut et al. 1998, ApJL, 496, L101
\xref Kuu97\_1630: Kuulkers, E., Parmar, A. N., Kitamoto, S., Cominsky, L. R., \& Sood, R. K. 1997, MNRAS, 291, 81
\xref Kuu98\_1630: Kuulkers, E., Wijnands, R., Belloni, T., Mendez, M., van der Klis, M., \& van Paradijs, J. 1998, ApJ, 494, 753
\xref Kuz99\_1758: Kuznetsov, S. I. et al. 1999, Astron. Letters, 25, 351
\xref Lev98\_2123 Levine, A., Swank, J., \& Smith, D.A. 1998, IAUC 6955
\xref Mai99\_1758: Main, D. S., Smith, D. M., Heindl, W. A., Swank, J., Leventhal, M., Mirabel, I. F., \& Rodriguez, L. F. 1999, ApJ, 525, 901
\xref Mar96\_1856: Marshall, F. E., Ebisawa, K., Remillard, R., \& Valinia, A. 1996, IAUC 6504
\xref Mar97\_1709 Marshall, F. E., Swank, J. H., Thomas, B., Angelini, L., Valinia, A., \& 
Ebisawa, K. 1997, IAUC 6543
\xref Mar97\_1737: Marshall, F. E., \& Smith, D. M. 1997, IAUC 6603
\xref Mar97\_1739--278: Marti, J., Mirabel, I. F., Duc, P.-A, Rodriguez, L. F. 1997 A\&A, 323, 
158

}
\end{multicols}

\noindent
\begin{tabular}{p{12.9cm}}
\hline
\\
\end{tabular}

\clearpage

\noindent
\begin{tabular}{p{12.9cm}}
\makebox[2.7cm][l]{}Table 2. (continued) Notes and References for ASM 
Transients\\
\\
\hline
\hline
\end{tabular}
\begin{multicols}{2}

{\footnotesize 
\xref Mar98\_1803: Marshall, F. E., Strohmayer, T., \& Remillard, R. 1998, IAUC 6891
\xref Mar98\_1858: Marshall, F. E. \& Chakrabarty, D., A\&A, 337, 815
\xref Mar98\_1758: Marti, J., Mereghetti, S., Chaty, S., Mirabel, I. F., Goldoni, P., \& Rodriguez, L. F. 1998, A\&A, 338, L95
\xref Mar99a\_1723: Marshall, F. E., \& Markwardt, C. B. 1999, IAU Circ., 7103
\xref Mar99b\_1723: Marshall, F.E., Ueda, Y. \& Markwardt, C. B. 1999, IAUC 7133
\xref Mar99\_1739--285: Markwardt, C. B., Marshall, F. E., Swank, J. H., \& Cui, W. 1999, IAUC 7300
\xref Mar99\_1859+226: Markwardt, C. B., Focke, W. B., Swank, J. H., \& Taam, R. E. 1999, BAAS, 31, 1555 
\xref Mar99\_1915: Markwardt, C. B., Swank, J. H., \& Taam, R. E. 1999, ApJL, 513, L37
\xref McC99\_1819: McCollough, M. L., Finger, M. H., \&Woods, P. M. 1999, IAUC 7257
\xref McC99\_1859+226: McCollough, M. L., \& Wilson, C. A. 1999 IAUC 7282
\xref Men98\_1655: Mendez, M., Belloni, T., \& van der Klis, M. 1998, ApJL, 499, L187
\xref Men98\_1608: Mendez, M., et al. 1998, ApJ, 494, L65
\xref Men99a\_1608 Mendez, M., van der Klis, M., Ford, E. C., Wijnands, R., \& van Paradijs, 
J. 1999 ApJL, 511, L49
\xref Men99b\_1608: Mendez, M., van der klis, M., Wijnands, R., Ford, E.C., van Paradijs, J., \& Vaughan, B.A. 1998, ApJ, 505, L23
\xref Mer97\_1758: Mereghetti, S., Cremonesi, D. I., Haardt, F., Murakami, T., Belloni, T., \&  Goldwurm, A. 1997, ApJ, 476, 829
\xref Mir94\_1915: Mirabel, I. F., \& Rodriguez, L. F. 1994, Nature, 371, 46
\xref Mir98\_1915: Mirabel, I. F., et al. 1998, A\&A, 330, L9
\xref Mor97\_1915: Morgan, E. H., Remillard, R., \& Greiner, J. 1997, ApJL, 482, L155
\xref Mot97\_0812: Motch, C., Haberl, F., Dennerl, K., Pakull, M.,\& Janot-Pacheco, E. 1997, 
A\&A, 323, 853
\xref Muk00\_2138: Mukerjee, K., Agrawal, P. C., Paul, B., Rao, A. R., Seetha, S., 
Kasturirangan, K. 2000, A\&A, 353, 239
\xref Mul98\_1803: Muller, J. M., et al. 1998, IAUC 6867
\xref Mun99\_1915: Muno, M.,  Morgan, E.,\& Remillard, R. 1999, ApJ, 527, 321
\xref Nat99\_1750: Natalucci, L., Cornelisse, R., Bazzano, A., Cocchi, M., Ubertini, P., Heise, 
J., In't Zand, J. J. M., \& 
Kuulkers, E. 1999, ApJL, 523, L45
\xref Nis99\_1744: Nishiuchi, M., et al. 1999, ApJ, 517, 436
\xref Oos98\_1630: Oosterbroek et al. 1998, A\&A, 340, 431
\xref Orl00\_0421: Orlandini, M., et al. 2000, A\&A (in press, astro-ph 0001530)
\xref Oro97a\_1655: Orosz, J. A., \& Bailyn, C. D. 1997, ApJ, 477, 876
\xref Oro 97b\_1655: Orosz, J. A., Remillard, R. A., Bailyn, C. D., \& McClintock, J. E. 1997, 
ApJL, 478, L83
\xref Orr98\_0421: Orr, A., et al. 1998, A\&A, 340, L190
\xref Pau98\_1858: Paul, B., \& Rao, A. R. 1998, A\&A, 337, 815
\xref Pir97\_1246: Piro, L., et al. 1997, IAUC 6538
\xref Poo97\_1915: Pooley, G. G., \& Fender, R. P. 1997, MNRAS, 292, 925
\xref Poo98\_2012: Pooley, G. G. 1998, IAUC 6926
\xref Poo99\_1859+226: Pooley, G. G., \& Hjellming, R. M. 1999, IAUC 7278
\xref Rei98a\_2030: Reig, P., \& Coe, M. J. 1998, MNRAS, 294, 118
\xref Rei98b\_2030: Reig, P., Stevens, J. B., \& Fabregat, J. 1998, MNRAS, 301, 42
\xref Rei99\_0812: Reig, P., \& Roche, P. 1999, MNRAS, 306, 95
\xref Rei99\_2030: Reig, P., \& Coe, M. 1999, MNRAS, 302, 700
\xref Rem97\_1354: Remillard, R., Marshall, F., \& Takeshima, T. 1997, IAUC 6772
\xref Rem97\_1755: Remillard, R., Levine, A., Swank, J., \& Strohmayer, T. 1997, IAUC 6710
\xref Rem98\_1711: Remillard, R. 1998, IAUC 6893
\xref Rem98\_1858: Remillard, R., \& Levine, A. 1998 IAUC 6826
\xref Rem98\_2012: Remillard, R., Levine, A., \& Wood, A. 1998, IAUC 6920

}
\end{multicols}

\noindent
\begin{tabular}{p{12.9cm}}
\hline
\\
\end{tabular}

\clearpage

\noindent
\begin{tabular}{p{12.9cm}}
\makebox[2.7cm][l]{}Table 2. (continued) Notes and References for ASM 
Transients\\
\\
\hline
\hline
\end{tabular}
\begin{multicols}{2}

{\footnotesize 
\xref Rem99\_1550: Remillard, R. A., McClintock, J. E., Sobczak, G. J., Bailyn, C. D., Orosz, J. A., Morgan, E. H., \& Levine, A. M. 1999, ApJL, 517, L127
\xref Rem99\_1630: Remillard, R., \& Morgan, E. 1999, BAAS, 31, 1421
\xref Rem99\_1655: Remillard, R. A., Morgan, E. M., McClintock, J. E., Bailyn, C. D., \& Orosz, J. A. 1999, ApJ, 522, 397
\xref Rem99\_1716: Remillard, R. A. 1999, Mem. Soc. Astron. Ital., 70, 881 (astro-ph 9805224)
\xref Rev98\_1755: Revnivtsev, M., Gilfanov M., \& Churazov, E. 1998, A\&A, 339, 483. See also Astron. Letters 25, 493 and ApJ, 511, 847
\xref Rev99\_0421: Revnivtsev, M. G., Emel'yanov, A. N., \& Borozdin, K. N. 1999, Astron. Letters, 25, 294
\xref Rev99\_1748: Revnivtsev, M. G., Trudolyubov, S. P., \& Borozdin, K. N., 2000, MNRAS, 312, 151
\xref Rev99\_1803: Revnivtsev, M., Borozdin, K., Emelyanov, A. 1999, A\&A, 344 L25
\xref Rev00\_1354: Revnivtsev, M., Borozdin, K., Priedhorsky, W. \& Vikhlinin, A. 2000, ApJ 530, 955
\xref Rey96\_2030: Reynolds, A. P., Parmar, A. N., Stollberg, M. T., Verbunt, F., Roche, P., Wilson, R. B., Finger, M. H. 1996, A\&A, 312, 872
\xref Roc98\_1808: Roche, P., Chakrabarty, D., Morales-Rueda, L., Hynes, R., Slivan, S. M., Simpson, C., \& Hewett, P. 1998, IAUC 6885
\xref Rut98\_1550: Rutledge, R., Fox, D., \& Smith, D. A. 1998, Astr. Tel. 36
\xref Rut99\_1608: Rutledge, R. E., Bildsten, L., Brown, E. F., Pavlov, G. G., \& Zavlin, V. E. 
1999, ApJ, 514, 945
\xref San99\_0115: Santangelo, A., et al. 1999, ApJL, 523, L85 
\xref San99\_1550: Sanchez-Fernandez, C., et al. 1999, A\&A, 348, L9
\xref Sha98\_1908: Shahbaz, T., Bandyopadhyay, R. M., Charles, P. A., Wagner, R. M., Muhli, 
P., Hakala, P., Casares, J., 
\& Greenhill, J. 1998, MNRAS, 300, 1035
\xref Sha99\_1655: Shabaz, T., van der Hooft, F., Casares, J., Charles, P. A., \& van Paradijs, J. 
1999 MNRAS, 306, 89
\xref Smi97\_1731: Smith, D. A., Morgan, E. H., \& Bradt, H. 1997, ApJL, 479, L137
\xref Smi97\_1758: Smith, D. M., Heindl, W. A., Swank, J., Leventhal, M., Mirabel, I. F., \& Rodriguez, L. F. 1997, ApJL, 489, L51
\xref Smi98\_1550: Smith, D. A. 1998, IAUC 7008
\xref Smi98\_1748: Smith, D.A., Levine, A., \& Wood, A. 1998 IAUC 6932
\xref Smi98\_1946: Smith, D. A., \& Takeshima, T. 1998 IAUC 7014
\xref Smi99\_1819: Smith, D.A., Levine, A. M., \& Morgan, E. H. 1999, IAUC 7253
\xref Smi99\_0421: Smith, D., Remillard, R., Swank, J., Takeshima, T., \& Smith, E. 1998, IAUC 
6855
\xref Sob99\_1550: Sobczak, G. J., McClintock, J. E., Remillard, R. A., Levine, A. M., Morgan, 
E. H., Bailyn, C. D., \& Orocz, J. E. 1999, ApJL, 517, L121
\xref Sob99\_1655: Sobczak, G. J., McClintock, J. E., Remillard, R. A., Bailyn, C. D., \& Orosz, 
J. A. 1999, ApJ, 520, 776
\xref Sof98\_1845: Soffitta, P., et al. 1998, ApJL, 494, L203
\xref Sor99\_2123: Soria, R., Wu, K., \& Galloway, D. 1999, MNRAS, 309, 528
\xref Ste97\_1145--619: Stevens, J. B., Reig, P., Coe, M. J., Buckley, D. A. H., Fabregat, J., \& 
Steele, I. A. 1997, MNRAS, 
288, 988
\xref Sto99\_2030: Stollberg, Mark T., Finger, Mark H., Wilson, R. B., Scott, D. M., Crary, D. 
J., \& Paciesas, W. S. 1999, 
ApJ, 512, 313
\xref Stu99\_1819: Stubbings, R., \& Pearce, A. 1999, IAUC 7253
\xref Tak97\_1843: Takeshima, T. 1997, IAUC 6595
\xref Tak98\_1858: Takeshima, T. Corbet, R., Marshall, F., Swank, J., \& Chakrabarty, D. 1998, 
IAUC 68
\xref Tin95\_1655: Tingay, S. J., et al. 1995, Nature, 374, 141
\xref Tom99\_1655: Tomsick , J. A., Kaaret, P., Kroeger R. A., \& Remillard, R. A. 1999, ApJ, 
512, 892
\xref Tom99\_2123: Tomsick, J. A., Halpern, J., Kemp, J., \& Kaaret, P. 1999, ApJ, 521, 341
\xref Tru99\_1737: Trudolyubov, S., et al. 1999, A\&A, 342, 496; see also IAUC 6599

}
\end{multicols}

\noindent
\begin{tabular}{p{12.9cm}}
\hline
\\
\end{tabular}

\clearpage

\noindent
\begin{tabular}{p{12.9cm}}
\makebox[2.7cm][l]{}Table 2. (continued) Notes and References for ASM 
Transients\\
\\
\hline
\hline
\end{tabular}
\begin{multicols}{2}

{\footnotesize 
\xref Ube97\_1826: Ubertini, P., Bazzano, A., Cocchi, M., Natalucci, L., Heise, J., Muller, J. M., \& In 't Zand, J. J. M. 1999, ApJL, 514, L27
\xref Ued97\_1737: Ueda et al. 1997, IAUC 6627
\xref Ued98\_0421: Ueda, Y., Ishida, M., Inoue, H., Dotani, T., Greiner, J., \& Lewin, W. H. G.. 1998, ApJ, 508, L167
\xref Uem99\_1859+226: Uemura, Kato, T., Pavlenko, E., Shugarov, S., \& Mitskevich, M., 1999 IAUC 7303
\xref Ung99\_0115:  Unger, S. J. Roche, P., Negueruela, I., Ringwald, F. A, Lloyd, C., \& Coe, M. J., et al. 1998, A\&A, 336, 960
\xref Var97\_1739--278: Vargas, M., et al. 1997, ApJL, 476, L23
\xref vdK00\_1808: van der Klis, M., Chakrabarty, D., Lee, J. C., Morgan, E. H.,Wijnands, R., Markwardt, C. B., \& Swank, J. H. 2000, IAUC 7358
\xref VaP96\_1908: vanParadijs, J. 1996, ApJL, 464, L1
\xref Vau98\_1808: Vaughan, B. A., et al. 1998, ApJL, 509, L145; also ApJL, 483, L115
\xref Wac00\_1808: Wachter, S., \& Hoard, D. W. 1998, IAUC 7363
\xref Wag99\_0421: Wagner, R.M., \& Starrfield, S.G. 1999, IAUC 6857
\xref Wen99\_1716: Wen, L., Levine, A., \& Bradt, H. 1999, BAAS 31, 1427
\xref Whi98\_2012: White, N. E., Ueda, Y., Dotani, T., \& Nagase, F. 1998, IAUC 6927
\xref Wij97\_1731: Wijnands, R. A. D., \&  van der Klis, M. 1997, ApJL, 482, L65
\xref Wij98a\_1808: Wijnand, R., \& van der Klis, M. 1998, Nature 294, 344
\xref Wij98b\_1808: Wijnands, R., \& van der Klis, M. 1998, ApJL, 507, L63
\xref Wij99\_1803: Wijnands, R., van der Klis, M. 1999, ApJ, 522, 965
\xref Wij99\_1819: Wijnands, R., \& van der Klis, M. 2000, ApJL, 528, L93
\xref Wil97\_1843: Wilson, R., Harmon, B., Scott, D., Finger, M., Robinsin, C., Chakrabarty, 
D., \& Prince, T. A. 1997, IAUC 
6586
\xref Wil98\_0111: Wilson, C. A., \& Finger, M. H. 1998, IAUC 7048
\xref Wil98\_1550: Wilson, C. A., Harmon, B. A., Paciesas, W. S., \& McCollough, M. L. 1998, 
IAUC 7010
\xref Wil99\_2138: Wilson, C. A., Finger, M. H., \& Scott, D. M. 1999, ApJ, 511, 367
\xref Woo99\_1859+226: Wood, A., Smith, D. A., Marshall, F. E., \& Swank, J. 1999, IAUC 
7274
\xref Woo99\_1744: Woods, P. M., et al. 1999, ApJ, 517, 431
\xref Yu97\_1608: Yu, W., et al. 1997, ApJL, 490, L153
\xref Yu99\_1908: Yu, W., Li, P., Zhang, W., \& Zhang. S. N. 1999, ApJL, 512, L35
\xref Zha96\_1608: Zhang, S. N., et al. 1996 A\&A Suppl. Ser., 120, 279
\xref Zha96\_1744: Zhang, W., Morgan, E. H., Jahoda, K., Swank, J. H., Strohmayer, T. E., 
Jernigan, J. G., \& Klein, R. I. 
1996 ApJL, 469, L2

}
\end{multicols}

\noindent
\begin{tabular}{p{12.9cm}}
\hline
\\
\end{tabular}

\begin{figure}
\centerline{\psfig{file=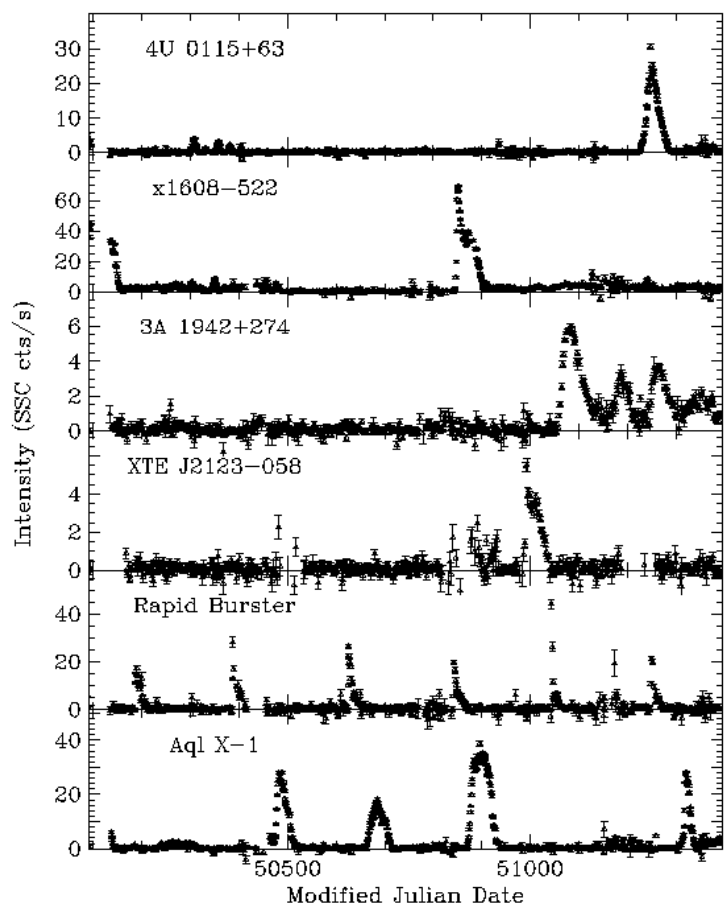, height=12cm, rheight=14cm}}
\caption[]{RXTE/ASM light curves for six neutron-star binary-system
transients for the period early January 1996 through mid August
1999. The data points represent 1-day averages of the 10 -- 20
(typical) daily measurements in the 1.5--12~keV band. 75~ct/s
corresponds to 1.0~Crab. MJD~50082 = 1996 Jan.~0.0.}
\end{figure}

\begin{figure}
\centerline{\psfig{file=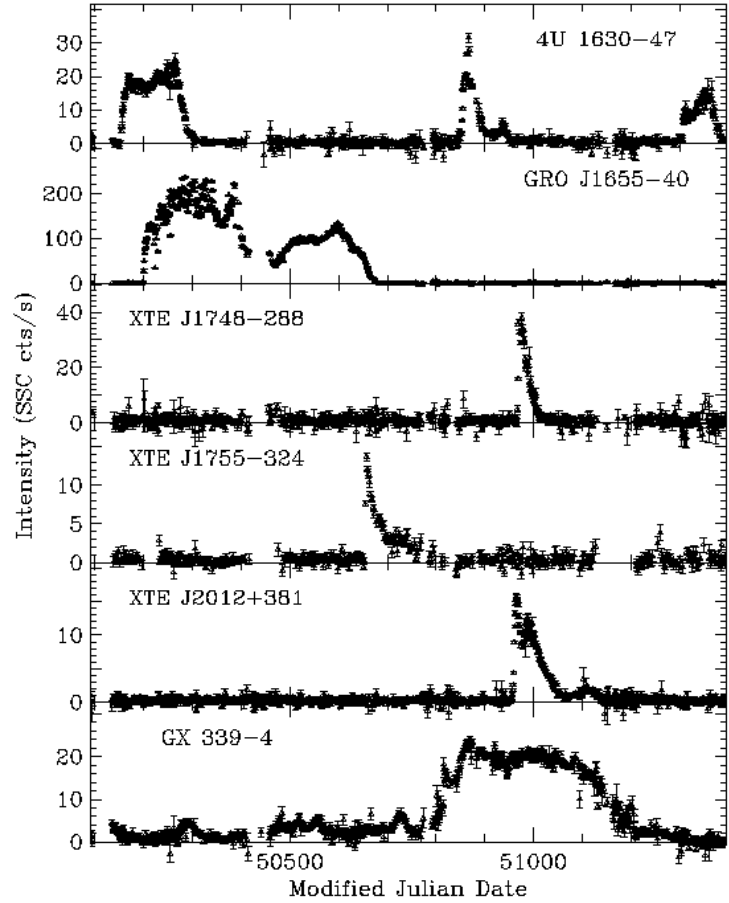, height=12cm, rheight=14cm}}
\caption[]{RXTE/ASM light curves for six black-hole binary-system
transients. GX 339--4 shows a transient bright soft state; it is not
listed as a transient in Table~1. See also caption to Fig.~1.}
\end{figure}

\end{document}